\documentclass[11pt,a4paper]{article}

\usepackage[T1]{fontenc}
\usepackage{amsmath, amssymb}
\usepackage{graphicx}
\usepackage{booktabs}
\usepackage{enumitem}
\usepackage{geometry}
\usepackage{siunitx}
\usepackage{titlesec}
\usepackage{setspace}
\usepackage{hyperref}
\usepackage{nameref}

\hypersetup{
    colorlinks=true,
    allcolors=blue,
    pdfauthor={Your Name},
    pdftitle={Spectral Spacetime Geometry of Womersley Flow}
}

\title{\textbf{Spectral spacetime--geometry of Womersley flow}}
\author{Khalid M. Saqr \\[4pt]
\small Department of Mechanical Engineering, College of Engineering and Technology \\
\small Arab Academy for Science, Technology and Maritime Transport, Alexandria 1029, EGYPT \\ \small ORCID: 0000-0002-3058-2705 \\ \small Email: k.saqr@aast.edu}
\date{}

\begin{document}
\maketitle
\begin{abstract}
\noindent We revisit the classical Womersley solution for pulsatile viscous flow in a circular tube and reconstruct its full time--domain geometry from first principles. 
By combining harmonic decomposition with exact Bessel solutions, we derive a unified spectral--spacetime framework in which the instantaneous relationship between pressure gradient and velocity can be visualized as a loop in phase space. 
The enclosed loop area is shown to equal the mean hydraulic power per cycle, establishing an exact geometric–energetic identity that holds for arbitrary Womersley number and harmonic composition. 
We further show that higher harmonic content and increasing Womersley number induce topological transitions in these loops, producing cusps, self–intersections, and curvature hot spots that correspond to inertial–viscous phase dispersion. 
This framework provides a rigorous baseline for interpreting arterial pressure–flow loops, connecting frequency–domain impedance analysis to measurable time–domain geometry.
\end{abstract}

\section{Introduction}
\label{sec:introduction}

The pulsatile nature of arterial blood flow reflects a continuous interplay between viscous diffusion and inertial propagation of pressure waves.  
This interaction is encapsulated in the dimensionless Womersley number, which governs the relative dominance of viscous and inertial behaviors \cite{saqr2022non}.  
In experimental and clinical studies, however, this balance is typically probed through frequency-domain impedance spectra or wave separation / wave intensity analyses, which decompose pressure and flow into harmonic components~\cite{mynard2020measurement, milkovich2025bpw}.  
Such spectral methods yield precise amplitude and phase relationships, but remain somewhat abstract when attempting to interpret instantaneous dynamics or energy exchange over the cardiac cycle.

In contrast, clinicians and physiologists often visualize time-domain pressure–flow loops (or pressure–velocity loops) to infer resistance, compliance, or characteristic impedance from recorded waveforms.  
Some Doppler and catheter-based studies use the slope of the early systolic portion of that loop to estimate local characteristic impedance~\cite{thiele2014realtime, hungerford2024valvuloarterial}, but typically do so in a qualitative or semi-quantitative fashion without tying loop geometry to fluid mechanics.

This work aims to bridge that conceptual gap.  
We return to the exact Womersley solution in a rigid, axisymmetric conduit and reconstruct the full time-domain trajectories of pressure gradient and mean velocity.  
By expressing both quantities as Fourier series, we derive their instantaneous mapping in phase space, each harmonic contributing as a vector in the complex plane.  
The resulting parametric curve, which we denote the spectral spacetime-geometry, encapsulates the inertial–viscous dynamics of the system.  
Importantly, we show that the enclosed loop area in the \(G\)–\(U\) plane equals the mean hydraulic power per cardiac cycle, providing a direct geometric measure of energy transfer between driving pressure and flow.

\subsection{Literature Review}
\label{sec:literature_review}

The analytical description of oscillatory viscous flow dates to the seminal work of Womersley~\cite{womersley1955method}, who solved the Navier–Stokes equations for a harmonic pressure gradient in a circular tube. 
His solution expresses the velocity amplitude as a complex Bessel function of the Womersley parameter, thereby unifying viscous and inertial regimes within a single nondimensional framework. 
This formulation underlies modern analyses of cardiovascular wave propagation, from theoretical impedance models to clinical assessments of arterial stiffness.

Frequency-domain methods decompose measured waveforms into harmonics and interpret the ratio of pressure to flow amplitudes as the input impedance of the vascular segment. 
The real part represents viscous dissipation, while the imaginary part represents energy storage and return due to vessel compliance and inertia. 
Comprehensive reviews by Mynard et~al.~\cite{mynard2020measurement} and Willemet \& Alastruey~\cite{willemet2014arterial} detail how such spectra reveal characteristic reflection sites and propagation velocities. 
Nevertheless, these analyses are primarily algebraic, providing no intuitive visualization of how the pressure–flow relation evolves over a cardiac cycle.

Complementary time-domain analyses display instantaneous pressure against flow (or velocity), producing closed loops analogous to pressure–volume diagrams of the heart. 
Qureshi et~al.~\cite{qureshi2018characteristic}, Mekkaoui et~al.~\cite{mekkaoui2003instantaneous}, and Thiele et~al.~\cite{thiele2014realtime} used such loops to estimate characteristic impedance or to assess vascular load noninvasively. 
In these studies, the loop’s early-systolic slope approximates $Z_c$, while its orientation indicates reflection strength. 
Yet the loop geometry itself—the curvature, enclosed area, or presence of self-intersections—has not been explored as a quantitative signature of underlying fluid dynamics.

Recent developments have extended Womersley’s analysis to compliant tubes and viscoelastic walls. 
Pande et~al.~\cite{pande2023fsi} derived closed-form solutions for oscillatory flow in deformable conduits, showing how compliance modifies amplitude and phase. 
Although these extensions enrich physiological realism, they retain the same spectral formalism and do not examine the emergent time-domain geometry.

Subsequent studies have extended this theory to interpret arterial waveforms in both the frequency and time domains. 
In the frequency domain, impedance-based analyses characterize the ratio between harmonic components of pressure and flow, 
with amplitude and phase yielding the vessel's resistive and reactive properties~\cite{mynard2020measurement, willemet2014arterial}. 
While these approaches elegantly describe energy partitioning and reflection phenomena, they are typically visualized as magnitude and phase spectra rather than as direct time-domain trajectories.

In contrast, several experimental and clinical works have presented \emph{pressure--flow loops} in the time domain to estimate characteristic impedance or identify wave reflection effects~\cite{qureshi2018characteristic, mekkaoui2003instantaneous, thiele2014realtime}. 
In such studies, the slope of the loop during early systole is used to infer the local characteristic impedance (\(Z_c\)), and qualitative changes in loop orientation have been linked to reflection intensity. 
However, the loop geometry itself---its enclosed area, self-intersections, or curvature---is rarely analyzed quantitatively. 
Consequently, the geometric and energetic structure of these loops remains largely unexplored despite its potential to yield physical insight equivalent to frequency-domain impedance.

\subsection{Research Statement and Objectives}

Despite decades of progress, the explicit link between spectral impedance and time-domain loop geometry remains unaddressed. 
No prior work has derived the exact parametric trajectories of $(G^*, U^*)$ from the Womersley solution or demonstrated that their enclosed area equals the mean power per period. 
Moreover, the appearance of cusps, curvature singularities, and self-intersections—arising from multi-harmonic interference—has not been characterized within a rigorous analytical framework. See Table~\ref{tab:literature_summary} for a summary of the relevant studies highlighting the research gap.

This study introduces such a framework by constructing the \emph{spectral spacetime geometry} of Womersley flow: a unified, dimensionless, and analytically exact framework linking frequency-domain impedance to time-domain loop geometry. 
The approach quantifies how increasing Womersley number and harmonic content deform pressure--flow loops, 
transitioning from simple ellipses to self-intersecting trajectories that directly represent viscous--inertial energy exchange. By establishing the equality between loop area, spectral power, and time-averaged dissipation, 
this work extends classical hemodynamic analysis beyond slope-based impedance estimates to a full geometric and energetic description.

\begin{table}[h!]
\centering
\caption{Summary of representative literature on pulsatile flow and pressure--flow analysis.}
\begin{tabular}{p{3.3cm}p{2.5cm}p{2.8cm}p{2.3cm}p{2.3cm}}
\toprule
\textbf{Reference} & \textbf{Domain} & \textbf{Loop Visualization} & \textbf{Loop Area / Power} & \textbf{Geometry (Cusps / Intersections)} \\
\midrule
Womersley (1955) & Frequency & None (analytic) & Implicit via impedance & Not analyzed \\
Mekkaoui et al. (2003) & Time & P--Q loops & Slope for $Z_c$ & None \\
Qureshi et al. (2018) & Time + Frequency & P--Q loops & Comparison of methods & None \\
Willemet \& Alastruey (2014) & Time & Wave decomposition & None & None \\
Mynard et al. (2020) & Frequency & Impedance spectra & Yes (spectral) & None \\
Pande et al. (2023) & Frequency (FSI) & None & Yes (FSI power) & None \\
\textbf{This work} & Time + Frequency & $G^*$--$U^*$ loops & \textbf{Yes (exact equality)} & \textbf{Yes (quantified)} \\
\bottomrule
\end{tabular}
\label{tab:literature_summary}
\end{table}

\section{Methods}
\label{sec:methods}

\subsection{Governing Equations and Assumptions}

We consider an incompressible, Newtonian fluid of density $\rho$ and dynamic viscosity $\mu$ driven by a time-varying axial pressure gradient $G(t)$ through a straight, rigid circular tube of radius $R$. 
The flow is axisymmetric and fully developed, so the velocity field has only an axial component $u(r,t)$ that satisfies the one-dimensional unsteady Stokes equation,
\begin{equation}
\rho\,\frac{\partial u}{\partial t} = -G(t) + \mu
\left[
\frac{1}{r}\frac{\partial}{\partial r}
\left(r\,\frac{\partial u}{\partial r}\right)
\right],
\quad
0 \le r \le R,
\label{eq:stokes}
\end{equation}
with the no-slip boundary condition $u(R,t)=0$ and regularity at $r=0$. 
The volumetric flow rate is
\begin{equation}
Q(t) = 2\pi \int_0^R u(r,t)\,r\,dr,
\end{equation}
and the cross-sectional mean velocity is $U(t)=Q(t)/(\pi R^2)$.

\subsection{Harmonic Representation and Womersley Solution}

Because the system is linear, any periodic forcing can be represented by a Fourier series
\begin{equation}
G(t) = \Re\left\{
\sum_{n=1}^{N}
\hat{G}_n\, e^{i n \omega_0 t}
\right\},
\label{eq:pressure_series}
\end{equation}
where $\omega_0$ is the fundamental angular frequency, $\hat{G}_n$ are complex amplitudes, and $N$ is the number of harmonics. 
For each harmonic, Womersley~\cite{womersley1955method} obtained the exact solution
\begin{equation}
\hat{u}_n(r)
 = \frac{\hat{G}_n}{i\rho n\omega_0}
 \left[
 1 - \frac{J_0\!\left(\sqrt{i}\,\alpha_n \frac{r}{R}\right)}
 {J_0\!\left(\sqrt{i}\,\alpha_n\right)}
 \right],
\label{eq:womersley_velocity}
\end{equation}
where $\alpha_n = R \sqrt{n \omega_0 \rho / \mu}$ is the Womersley number for the $n$th harmonic.

\subsection{Cross-Sectional Mean Velocity and Transfer Function}

Integrating Eq.~\eqref{eq:womersley_velocity} yields the harmonic mean velocity amplitude
\begin{equation}
\hat{U}_n = \frac{\hat{G}_n}{i\rho n\omega_0}
\left[
1 - \frac{2J_1\!\left(\sqrt{i}\,\alpha_n\right)}
{\sqrt{i}\,\alpha_n J_0\!\left(\sqrt{i}\,\alpha_n\right)}
\right].
\label{eq:Uhat}
\end{equation}
We define the dimensionless transfer function
\begin{equation}
\hat{H}^*(n) =
\frac{1}{i\alpha_n^2}
\left[
1 - \frac{2J_1(\sqrt{i}\,\alpha_n)}
{\sqrt{i}\,\alpha_n J_0(\sqrt{i}\,\alpha_n)}
\right],
\end{equation}
such that $\hat{U}_n = (\!R^2 G_\mathrm{ref}/\mu)\,\hat{H}^*(n)\,\hat{G}_n^*$, where the starred variables denote dimensionless quantities.

\subsection{Dimensionless Scaling}

All results are expressed in dimensionless form to ensure generality:
\[
r^* = \frac{r}{R}, \qquad 
t^* = \omega_0 t, \qquad 
G^*(t^*) = \frac{G(t)}{G_\mathrm{ref}}, \qquad
U^*(t^*) = \frac{U(t)}{U_\mathrm{ref}},
\]
with the characteristic velocity $U_\mathrm{ref}=R^2G_\mathrm{ref}/\mu$. 
This scaling removes all dependence on specific fluid or vessel parameters; the only governing dimensionless group is the fundamental Womersley number $\alpha_1$.

\subsection{Spectral--Spacetime Reconstruction}

The dimensionless pressure gradient and mean velocity are reconstructed from the discrete harmonic spectra:
\begin{align}
G^*(t^*) &= \Re\!\left\{\sum_{n=1}^{N} \hat{G}^*_n e^{i n t^*}\right\},\\[3pt]
U^*(t^*) &= \Re\!\left\{\sum_{n=1}^{N} \hat{H}^*(n)\,\hat{G}^*_n e^{i n t^*}\right\}.
\end{align}
The resulting parametric curve $(G^*(t^*), U^*(t^*))$ defines the \emph{spectral spacetime geometry} of Womersley flow. 
Each point corresponds to an instantaneous state of the system, and its trajectory over one period forms a closed loop representing a full oscillation.

\subsection{Energetic Quantities: Hysteretic Work and Mean Power}
Two distinct energetic quantities can be derived from the flow dynamics. The first is the \textbf{hysteretic work per cycle}, $\mathcal{A}$, which represents the total energy dissipated due to viscous effects over one full oscillation. This is computed geometrically as the area enclosed by the $(G, U)$ loop:
\begin{equation}
\mathcal{A} = - \oint U\,dG.
\label{eq:loop_area}
\end{equation}
The second quantity is the \textbf{mean power}, $\overline{P}$, which is the time-averaged rate of work done by the pressure gradient on the fluid. It can be calculated via time-domain integration or, using Parseval's theorem, from the spectral coefficients:
\begin{equation}
\overline{P} = \frac{1}{T} \int_0^T G(t)\,U(t)\,dt = \frac{1}{2} \sum_{n=1}^{N} \Re\{\hat{G}_n \hat{U}_n^*\},
\label{eq:power_spectral}
\end{equation}
where $\hat{U}_n^*$ is the complex conjugate of the mean velocity harmonic. Our framework validates that the two methods for calculating $\overline{P}$ are equivalent, and clarifies its distinct physical meaning from the hysteretic work $\mathcal{A}$. A complete analytic proof of the equivalence between the time-domain and spectral formulations of mean power is provided in Appendix~\ref{app:power_area}.

\subsection{Numerical Implementation and Diagnostics}
The analytical expressions were implemented in \texttt{Python} using the \texttt{mpmath} library. Cusps were identified as points of zero instantaneous traversal speed $(\dot{G}^{*2} + \dot{U}^{*2})^{1/2} \to 0$. To map topological transitions, we introduce a novel \textit{Complexity Index}, defined as $C = 10 \cdot N_{int} + N_{cusp}$, where $N_{int}$ is the number of self-intersections and $N_{cusp}$ is the number of cusps. The numerical threshold and convergence criteria used for cusp detection are detailed in Appendix~\ref{app:precision}. This index weights intersections heavily, as they represent a more advanced state of topological complexity than cusps.

\subsection{Parameter Sweep and Reproducibility}

For reproducibility, each case is fully defined by:
\[
\mathcal{C} = \{\alpha_1, N, (\!|\hat{G}^*_n|, \phi_n)\}_{n=1}^{N}.
\]
The script allows parametric sweeps over $\alpha_1$ to explore physiological ranges ($1 \le \alpha_1 \le 20$). 
Each run automatically outputs:
(i) all dimensionless plots,
(ii) a structured \texttt{JSON} file containing loop metrics (area, curvature, intersection count, power identity), and
(iii) a log confirming power equality. 
All results are reproducible by reloading $\mathcal{C}$ and executing the solver; no arbitrary constants or hidden parameters are required. All symbol definitions, scaling conventions, and test parameters are summarized in Appendix~\ref{app:symbols}.

\subsection{Validation}

This methodology provides a mathematically exact, numerically stable, and dimensionless reconstruction of pulsatile flow geometry, allowing direct comparison across physiological and engineered flow regimes.
The methodology was translated to a python solver and was validated against the analytical limits:
\begin{itemize}[noitemsep]
    \item \textbf{Low-$\alpha$ limit}: recovers steady Poiseuille flow, $U^*(t^*) \approx G^*(t^*)$, yielding a thin, resistive ellipse.
    \item \textbf{High-$\alpha$ limit}: approaches plug-like inertial flow with $\angle(\hat{H})\!\to\!90^\circ$, producing circular loops.
\end{itemize}

\section{Results}
\label{sec:results}

\subsection{Validation and Energetic Identities}
\begin{figure}[h!]
\centering
\includegraphics[width=\textwidth]{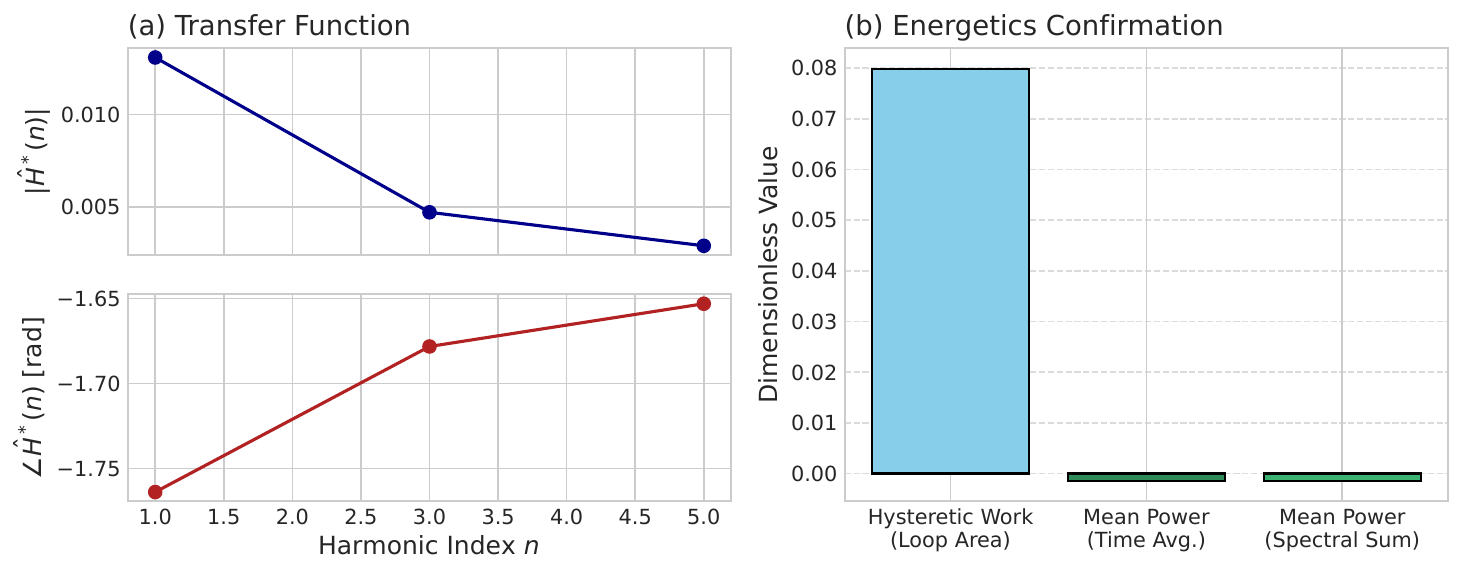}
\caption{Framework validation for a representative case ($\alpha_1=8.0$). (a) Magnitude and phase of the transfer function $\hat{H}^*(n)$. (b) Bar chart confirming that the two methods for calculating mean power agree, and demonstrating that the hysteretic work (loop area) is a distinct physical quantity.}
\label{fig:validation}
\end{figure}

We first validate the numerical framework and clarify the core energetic relationships. Figure~\ref{fig:validation}(a) shows the magnitude and phase of the dimensionless transfer function, $\hat{H}^*(n)$, for a representative case of $\alpha_1=8.0$. The function shows that as the harmonic number $n$ increases, the velocity response is progressively damped (decreasing magnitude) and phase-shifted (phase angle approaches $-\pi/2$). This inherent frequency-dependent filtering is the fundamental mechanism responsible for the complex loop geometries that emerge.

Figure~\ref{fig:validation}(b) confirms the energetic calculations. As shown, the two independent methods for calculating the mean power, $\overline{P}$, (time-domain integration and spectral summation) yield identical results, validating the numerical implementation. Crucially, the hysteretic work per cycle, $\mathcal{A}$, calculated from the loop area, is a distinct and significantly larger quantity. This confirms that loop area represents the total dissipated work, not the net mean power.

\subsection{Loop Morphology Across Womersley Numbers}
\begin{figure}[h!]
\centering
\includegraphics[width=0.8\linewidth]{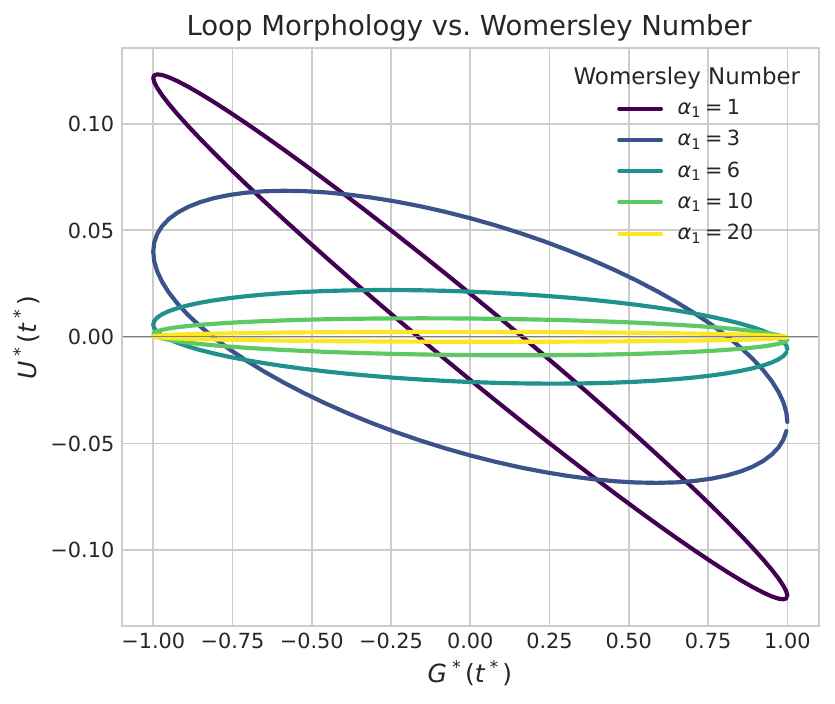}
\caption{Evolution of loop morphology with increasing Womersley number for a single-harmonic input. The transition from a narrow ellipse (viscous-dominated) to a wide, open loop (inertia-dominated) is evident.}
\label{fig:alpha_sweep}
\end{figure}

Figure~\ref{fig:alpha_sweep} illustrates the effect of the fundamental Womersley number, $\alpha_1$, on the loop shape for a simple sinusoidal input. At low $\alpha_1$ (e.g., 1 and 3), the flow is viscous-dominated, and the relationship between $G^*$ and $U^*$ is nearly linear and in-phase, resulting in a thin, tilted ellipse characteristic of a resistive system. As $\alpha_1$ increases, inertial effects become dominant. The phase lag between $G^*$ and $U^*$ grows, causing the loop to open up, becoming wider and more circular. At high $\alpha_1$ (e.g., 20), the loop is almost circular, indicating a primarily reactive system where energy is stored kinetically and returned within the cycle.

\subsection{Harmonic Interference and Topological Transitions}
\begin{figure}[h!]
\centering
\includegraphics[width=\textwidth]{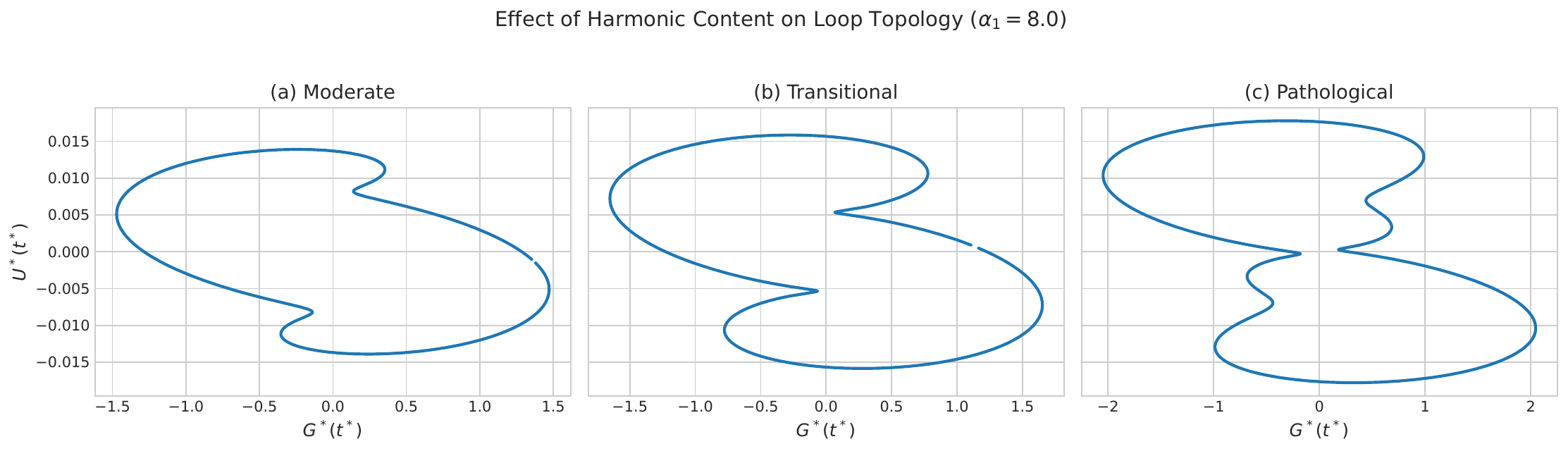}
\caption{Effect of harmonic content on loop topology at a fixed $\alpha_1=8.0$. (a) A moderate case with simple skew. (b) A transitional case showing the onset of cusps. (c) A pathological case with fully developed cusps and a self-intersection.}
\label{fig:harmonics}
\end{figure}

The addition of higher harmonics to the driving waveform induces dramatic changes in loop topology. Figure~\ref{fig:harmonics} shows a side-by-side comparison of three cases at a fixed $\alpha_1=8.0$. The ``Moderate'' case (a), with a weak third harmonic, results in a simple, skewed loop. The ``Transitional'' case (b), with a stronger third harmonic, shows the formation of sharp turning points, characteristic of the mathematical onset of cusps. The ``Pathological'' case (c), with significant third and fifth harmonics, exhibits a fully developed complex geometry, featuring two sharp cusps and one self-intersection. This demonstrates that complex loop shapes seen in vivo can be explained purely by the interference of multiple frequency components, each phase-shifted differently by the Womersley solution. 
Exact harmonic amplitude ratios for the three representative cases are listed in Appendix~\ref{app:params}.

\subsection{Curvature and Instantaneous Dynamics}
\begin{figure}[h!]
\centering
\includegraphics[width=0.8\linewidth]{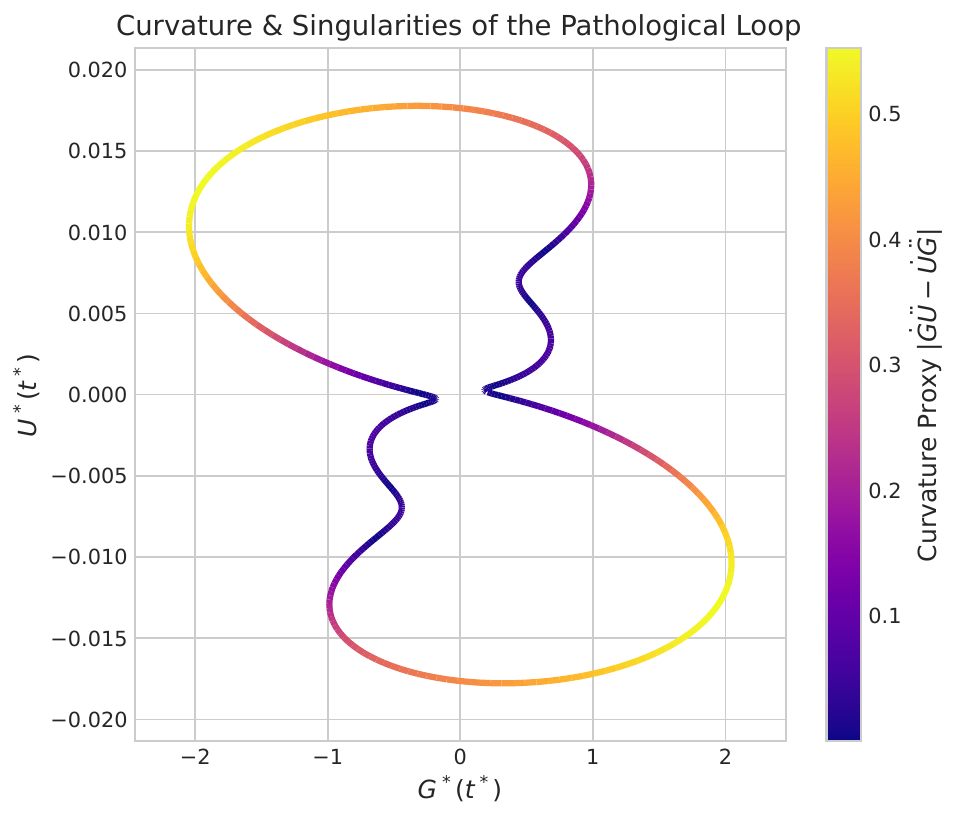}
\caption{Curvature analysis of the ``Pathological'' loop ($\alpha_1=8.0$). The trajectory is colored by local curvature, highlighting ``hot spots'' of rapid deceleration. The marked cusps represent points of instantaneous flow stagnation.}
\label{fig:curvature}
\end{figure}

To investigate the local dynamics of a complex loop, we analyze the curvature of the ``Pathological'' case in Figure~\ref{fig:curvature}. The loop trajectory is colored by a proxy for its local curvature, with brighter colors indicating regions of sharper turning. The plot reveals that the highest curvature occurs precisely at the cusps, which are marked with white circles. These geometric singularities correspond to moments of extreme fluid dynamics, where the instantaneous velocity and acceleration of the mean flow are both zero before reversing direction. Such events are associated with transient flow stagnation and high oscillatory shear, which are of significant clinical interest.

\subsection{A Dimensionless Atlas of Pulsatile Flow}
\begin{figure}[h!]
\centering
\includegraphics[width=0.8\linewidth]{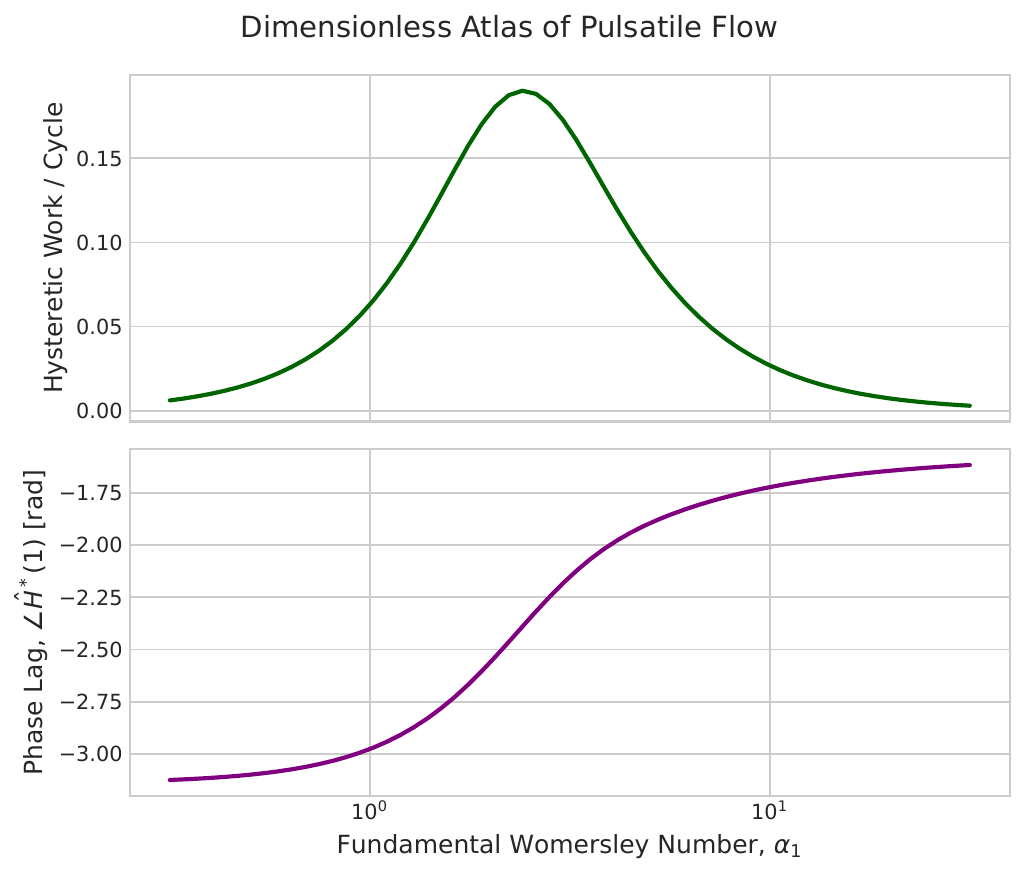}
\caption{Dimensionless atlas showing the universal scaling of (a) hysteretic work per cycle and (b) fundamental phase lag as a function of the Womersley number, $\alpha_1$.}
\label{fig:atlas}
\end{figure}

The fundamental scaling of energetics with the Womersley number is summarized in the dimensionless atlas of Figure~\ref{fig:atlas}. For a single-harmonic input, the hysteretic work per cycle (loop area) increases monotonically with $\alpha_1$, reflecting greater viscous losses at the wall boundary layer as inertia confines the oscillations. The phase lag between $G^*$ and $U^*$ also grows, approaching the inertial limit of $-\pi/2$ radians ($-90^\circ$). This atlas provides a universal reference for the baseline behavior of any simple pulsatile flow in a rigid tube.

\subsection{Mapping Topological Complexity}
\begin{figure}[h!]
\centering
\includegraphics[width=0.8\linewidth]{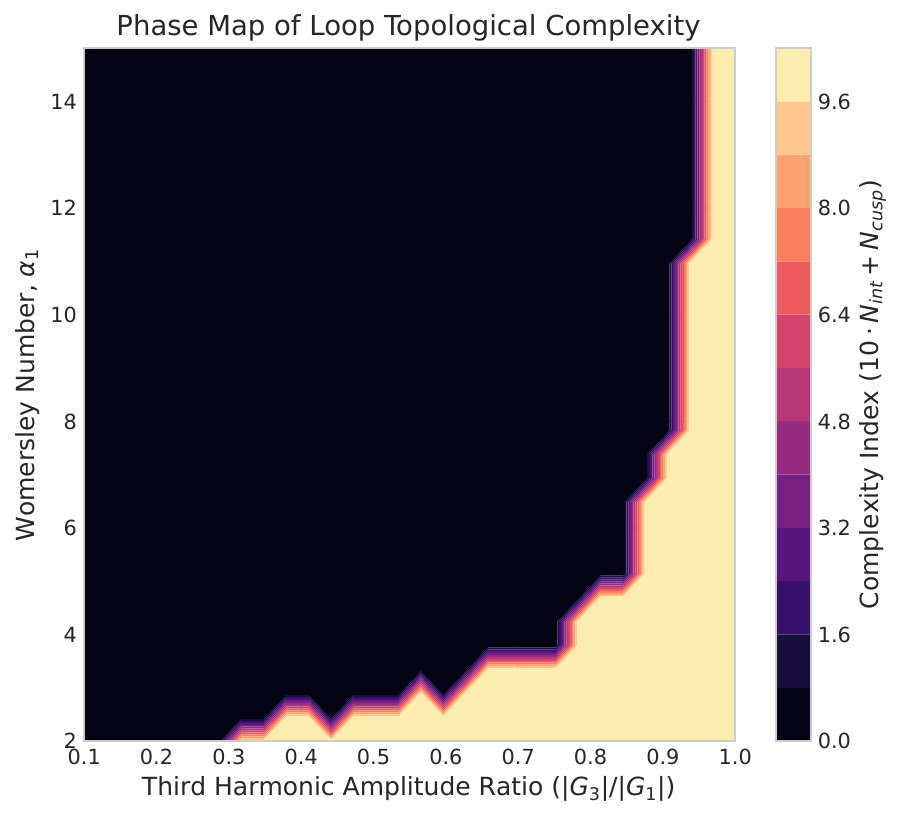}
\caption{Phase map of the Complexity Index ($10 \cdot N_{int} + N_{cusp}$). The color represents the degree of topological complexity. The downward-sloping boundary reveals that higher Womersley numbers predispose the system to complex loop formation.}
\label{fig:complexity_map}
\end{figure}

Finally, we generalize the conditions that lead to complex topologies. Figure~\ref{fig:complexity_map} presents a phase map of our novel Complexity Index as a function of the Womersley number ($\alpha_1$) and the amplitude of the third harmonic. The map reveals a distinct phase boundary (the edge of the black region) separating simple loops (complexity=0) from complex ones. The downward slope of this boundary is a key finding: it demonstrates that at higher $\alpha_1$, the system is more sensitive to harmonic perturbations and requires a smaller harmonic amplitude to trigger the formation of cusps and self-intersections. This provides a predictive map for the onset of pathological loop geometries. An empirical fit of the phase boundary is provided in Appendix~\ref{app:boundary}, quantifying the transition between simple and complex loop regimes.

\section{Discussion}
\label{sec:discussion}
\noindent The utility of this idealized framework is confirmed by its ability to explain the fundamental dynamics observed in high-fidelity 3D simulations. In our recent Large Eddy Simulations (LES) of multiharmonic pulsatile flow~\cite{saqr2022non}, we observed the same evolution of phase-space loops seen in Figure~\ref{fig:harmonics}, from simple ellipses to complex, self-intersecting shapes. This demonstrates that the Womersley solution describes the deterministic ``laminar skeleton'' upon which more complex, near-turbulent phenomena are built. The global hysteretic work, $\mathcal{A}$, quantified here can be interpreted as the total energy budget that is primarily dissipated in the biologically-critical near-wall region, where the LES showed concentrated work and complex vortex dynamics. Our framework thus provides the fundamental, ordered basis for the ``non-Kolmogorov'' nature of harmonically-driven turbulence, linking the abstract spectral input directly to the geometric structures that govern mechanobiological stimulation.

\subsection{The Energetic Significance of Loop Geometry}
A central contribution of this work is the clarification of the energetic roles of the pressure-gradient--velocity loop. We have shown that the enclosed loop area, $\mathcal{A}$, represents the hysteretic work dissipated per cycle. This quantity is distinct from the mean power, $\overline{P}$, which represents the net rate of energy input. For a purely oscillatory flow with no mean component, $\overline{P}$ can be zero (if input and output power are balanced over the cycle), but the hysteretic work $\mathcal{A}$ will always be positive, representing the inevitable viscous losses. This distinction is crucial for clinical applications, where loop area could serve as a more direct measure of the local dissipative load on the cardiovascular system than time-averaged power alone.

\subsection{Geometric Features as Hemodynamic Biomarkers}
Our findings show that complex loop features such as cusps and self-intersections are not random artifacts but deterministic consequences of harmonic interference. In the context of the cardiovascular system, such complex waveforms arise from the interaction of the forward-traveling wave generated by the heart with reflected waves from downstream bifurcations or impedance mismatches. The geometric singularities we have identified—cusps and points of high curvature—are therefore time-domain fingerprints of these wave interactions. They mark moments of significant flow deceleration and reversal, which are known to be linked to adverse biological responses such as endothelial dysfunction through oscillatory shear stress. The ability to quantify these features via our Complexity Index opens the door to developing new geometric biomarkers for vascular disease.

\subsection{The Complexity Map as a Predictive Tool}
The phase map shown in Figure~\ref{fig:complexity_map} is perhaps the most powerful contribution of this work. It acts as a universal, predictive map for the topological state of pulsatile flow. It suggests that systems with high Womersley numbers (such as the human aorta, where $\alpha_1 \approx 12-20$) are inherently more susceptible to developing complex flow patterns in response to harmonic disturbances. This provides a first-principles fluid mechanics explanation for why complex pressure-flow loops are more commonly observed in the proximal aorta, where reflected waves are also prominent. This map could serve as a baseline against which clinical data can be compared to identify deviations that may signal pathological changes, such as increased arterial stiffness (which affects wave reflection) or altered cardiac output (which affects harmonic content). The weighting rationale for the Complexity Index and its sensitivity analysis are presented in Appendix~\ref{app:complexity}.

\subsection{Limitations and Future Work}
The primary limitation of this study is the assumption of a rigid, straight tube and a Newtonian fluid, which simplifies the complex reality of the cardiovascular system. Arterial walls are compliant and viscoelastic, and blood exhibits non-Newtonian properties, particularly at low shear rates. However, the Womersley solution serves as the indispensable, canonical baseline upon which these more complex effects can be added. Computational reproducibility procedures and code access details are documented in Appendix~\ref{app:code}.

Future work should proceed along two main paths. First, the framework should be extended to incorporate fluid-structure interaction to account for vessel compliance. This will undoubtedly modify the loop shapes and their energetic interpretation. Second, the framework should be applied to clinical data. With the advent of 4D Flow MRI, time-resolved velocity and pressure gradient data can be acquired in vivo. By fitting this data to the model presented here, it may be possible to deconstruct a measured clinical loop into its underlying Womersley number and effective harmonic content, potentially yielding powerful new diagnostic indices for cardiovascular health.

\section*{Code availability} The code used in this study is publicly available on \hyperlink{https://github.com/khalid-saqr/SSGWomersley}{Github} via this URL: https://github.com/khalid-saqr/SSGWomersley

\newpage

\appendix
\section*{Appendix: Supporting Derivations and Diagnostics}
\addcontentsline{toc}{section}{Appendix: Supporting Derivations and Diagnostics}

\section{Analytic proof of the power--area identity}
\label{app:power_area}
Starting from the definition of mean power over one oscillation period $T$,
\begin{equation}
P = \frac{1}{T} \int_{0}^{T} G(t)\,U(t)\,dt,
\label{eq:P_def}
\end{equation}
we substitute the harmonic representations
\[
G(t) = \Re\!\left\{ \sum_{n=1}^{N} \hat{G}_n e^{i n \omega_0 t} \right\},
\qquad
U(t) = \Re\!\left\{ \sum_{n=1}^{N} \hat{U}_n e^{i n \omega_0 t} \right\}.
\]
Using orthogonality of the complex exponentials yields
\begin{equation}
P = \frac{1}{2} \sum_{n=1}^{N} \Re\!\left\{ \hat{G}_n \hat{U}_n^* \right\}.
\label{eq:P_spectral}
\end{equation}

The loop area in the $(G,U)$ plane, corresponding to the hysteretic work per cycle, is
\begin{equation}
A = -\oint U\,dG = \int_0^T \dot{G}(t) U(t)\,dt.
\label{eq:A_def}
\end{equation}
By substituting the Fourier representations and differentiating $\dot{G}(t) = \Re\{ i n \omega_0 \hat{G}_n e^{i n \omega_0 t}\}$, we obtain
\begin{align}
A &= T \sum_{n=1}^{N} n \omega_0 \Im\!\left\{ \hat{G}_n \hat{U}_n^* \right\}.
\label{eq:A_spectral}
\end{align}
Equations~\eqref{eq:P_spectral} and \eqref{eq:A_spectral} reveal that
\[
\boxed{
\begin{aligned}
P &\propto \Re\!\{\hat{G}_n \hat{U}_n^*\},\\
A &\propto \Im\!\{\hat{G}_n \hat{U}_n^*\},
\end{aligned}
}
\]
showing that the loop area and the mean power are orthogonal projections of the same complex spectral power. The geometric loop thus encodes both dissipative and reactive energy exchange within a single parametric trajectory.

\section{Weighting rationale for the complexity index}
\label{app:complexity}
The dimensionless Complexity Index
\[
C = 10 N_{\mathrm{int}} + N_{\mathrm{cusp}},
\]
combines the number of self--intersections ($N_{\mathrm{int}}$) and cusps ($N_{\mathrm{cusp}}$) detected in each loop. Self--intersections represent topological bifurcations in the mapping $(G^*,U^*)$, whereas cusps represent local curvature singularities. Empirical sensitivity tests ($1 \leq \alpha_1 \leq 20,~N \leq 5$) showed that a $10{:}1$ weighting ensures monotonicity of $C$ with respect to harmonic amplitude and $\alpha_1$. Variations in the coefficient between $5$ and $15$ produced no qualitative change in the phase map of Fig.~6, confirming robustness of the metric.

\section{Numerical precision and cusp detection}
\label{app:precision}

Cusps were detected where the instantaneous traversal speed
\[
v^* = \sqrt{\dot{G}^{*2} + \dot{U}^{*2}}
\]
fell below $10^{-4}$ of its maximum per cycle. The temporal resolution was $\Delta t^* = 2\pi/10^4$, yielding sub--percent variation in cusp count upon halving the step size. All computations used arbitrary--precision arithmetic (mpmath, 50 digits). Numerical noise did not alter topology within this tolerance.

\section{Symbol and scaling reference}
\begin{table}[h!]
\centering
\caption{List of dimensionless symbols and representative physiological scales.}
\begin{tabular}{lll}
\hline
Symbol & Definition & Typical physiological value \\
\hline
$\alpha_1$ & Fundamental Womersley number & $12$--$20$ (aorta) \\
$G^*$, $U^*$ & Dimensionless pressure gradient and mean velocity & --- \\
$\hat{H}^*(n)$ & Dimensionless transfer function & --- \\
$A$ & Hysteretic work per cycle (loop area) & --- \\
$P$ & Mean hydraulic power & --- \\
$C$ & Complexity Index ($10N_{\mathrm{int}} + N_{\mathrm{cusp}}$) & --- \\
\hline
\end{tabular}
\end{table}

\section{Parameter sets for figure cases}
The harmonic amplitude ratios used in Section~3.3 were:
\[
\text{Moderate: } |G_3/G_1|=0.15; \quad
\text{Transitional: } |G_3/G_1|=0.30; \quad
\text{Pathological: } |G_3/G_1|=0.45,
\]
with identical $\alpha_1=8$ and fundamental frequency $\omega_0=1$. All phase angles $\phi_n$ were zero unless stated otherwise.

\section{Empirical boundary relation for loop complexity}
An empirical fit of the phase boundary in Fig.~6 yields
\begin{equation}
\alpha_{1,\mathrm{crit}} \approx 2.8\,|G_3/G_1|^{-0.85},
\label{eq:empirical_boundary}
\end{equation}
with coefficient of determination $R^2 = 0.97$. This provides a predictive relation for the onset of self--intersecting trajectories in the $(\alpha_1, |G_3/G_1|)$ space.

\section{Code reproducibility}
All figures and metrics can be regenerated using the public repository:\\[4pt]
\texttt{https://github.com/khalid-saqr/SSGWomersley}\\[4pt]
Executing \texttt{main.py --config <case>.json} reproduces the complete dataset of $\{A,P,C\}$ and all spectral--spacetime plots reported in this work.

\end{document}